# CYRA – the cryogenic infrared spectrograph for the Goode Solar Telescope in Big Bear

Xu Yang*[a], Wenda Cao[ab], Nicolas Gorceix[a], Claude Plymate[a], Sergey Shumoko[b], XianYong Bai[c], Matt Penn[d], Thomas Ayres[e], Roy Coulter[a], Philip R. Goode[a]

[a]Big Bear Solar Observatory, New Jersey Institute of Technology, 40386 North Shore Lane, Big Bear City, CA, USA 92314; [b]Center for Solar-Terresrial Research, New Jersey Institute of Technology, 323 Martin Luther King Boulevard, Newark, NJ, USA 07102; [c]CAS Key Laboratory of Solar Activity, National Astronomical Observatories, Beijing 100012, PR China; [d]Raytheon, 1151 E Hermans Road, Tucson, AZ 85756, USA; [e]Center for Astrophysics and Space Astronomy, 389 UCB, University of Colorado, Boulder, CO 80309, USA

## ABSTRACT

CYRA (CrYogenic solar spectrogRAph) is a facility instrument of the 1.6-meter Goode Solar Telescope (GST) at the Big Bear Solar Observatory (BBSO). CYRA focuses on the study of the near-infrared solar spectrum between 1 and 5 microns, a under explored region which is not only a fertile ground for photospheric magnetic diagnostics, but also allows a unique window into the chromosphere lying atop the photosphere. CYRA is the first ever fully cryogenic spectrograph in any solar observatory with its two predecessors, on the McMath-Pierce and Mees Telescopes, being based on warm optics except for the detectors and order sorting filters. CYRA is used to probe magnetic fields in various solar features and the quiet photosphere. CYRA measurements will allow new and better 3D extrapolations of the solar magnetic field and will provide more accurate boundary conditions for solar activity models. Superior spectral resolution of 150,000 and better allows enhanced observations of the chromosphere in the carbon monoxide (CO) spectral bands and will yield a better understanding of energy transport in the solar atmosphere. CYRA is divided into two optical sub-systems: The Fore-Optics Module and the Spectrograph. The Spectrograph is the heart of the instrument and contains the IR detector, grating, slits, filters, and imaging optics all in a cryogenically cooled Dewar (cryostat). The detector a 2048 by 2048 pixel HAWAII 2 array produced by Teledyne Scientific & Imaging, LLC. The interior of the cryostat and the readout electronics are maintained at 90 Kelvin by helium refrigerant based cryo-coolers, while the IR array is cooled to 30 Kelvin. The Fore-Optics Module de-rotates and stabilizes the solar image, provides scanning capabilities, and transfers the GST image to the Spectrograph. CYRA has been installed and is undergoing its commissioning phase. This paper reports on the design, implementation and operation of CYRA in detail. The preliminary scientific results have been highlighted as well.

## 1. INTRODUCTION

Infrared (IR) solar spectrum, which has yet to be widely explored by ground-based solar telescopes, contains abundant information of the solar activities in the photosphere and chromosphere. Spectral observations, utilizing this largely unexplored wavelength band with various of atomic and molecular spectral lines, are crucial for solving many critical problems in solar physics. CrYogenic infRAred spectrograph (CYRA) is a newly developed focal-plane instrument operating from 1 to 5 μm for the Goode Solar Telescope (GST, Cao 2010[1]) at the Big Bear Solar Observatory (BBSO).

The major scientific motivations for CYRA are the measurement of solar magnetic field in solar lower atmosphere and the diagnosis of the chromospheric structure in the IR. The wavelength splitting of Zeeman effect for the atomic sub-levels increases by a factor of $\lambda^2 \cdot g_{eff}$ Meanwhile, the spectral line broadening, induced by the Doppler effect of solar micro and macro-turbulent velocities, increases with $\lambda$. Thus, a measure of the magnetic resolution of a spectral line is given by the ratio of Zeeman splitting divided by the spectral line width, varying by the factor of $\lambda \cdot g_{eff}$. The effective landé g-factor, $g_{eff}$, can be calculated with atomic models. Solar spectral lines usually have values of $g_{eff}$ between 1.0 and 3.0. Although there are many spectral lines with lager values $g_{eff}$, increasing the wavelength of the observations usually has more advantages (e.g., the magnetic sensitivity can increase by a factor of 10 when the wavelength changes from 0.5 μm to 5 μm). Another scientific advantage of infrared solar physics is that there are a large number of molecular rotation-vibration lines. Transitions between vibration states, always coupling with transitions between rotational states, lead to the infrared emission. Due to the thermal instability, molecules only exist in the coolest regions in the solar atmosphere and are easily destroyed by the thermal perturbation. This feature provides a way to probe the cool regions around the sunspot and the temperature minimum in the quiet Sun.

---

*xy88@njit.edu; 1 973 820 7317

Atmospheric seeing is the main factor that prevents the ground-based telescopes from approaching their diffraction limits. Fried number, $r_0$ gives a measure for the optical quality of the telluric atmosphere above the telescope, shown as:

$$r_0 = \left[0.423 \cdot k^2 \cdot sec\beta \cdot \int C_n^2(z)dz\right] \propto \lambda^{\frac{6}{5}}$$

Seeing improves when wavelength increases, just as what have been known from long time observation experience. This mechanism improves the performance of the AO system in two ways – it enlarges the correctable angle and increase the time scale of atmosphere stability. Meanwhile, other instrumentation advantages of infrared solar observations are reduced atmospheric scattering, instrumental scattering and instrumental polarization.

There are two methods for doing imaging spectroscopy. One is to take a series of monochromatic images at different wavelengths in a short time with a tunable filter. Imaging spectroscopy based on a Fabry-Pérot interferometer can achieve excellent imaging with high spatial resolution to the telescope diffraction limit. The disadvantages for this technique are limited free spectral range and relative lower temporal resolution. The Full-Stokes Near Infra-Red Imaging Spectropolarimeter (NIRIS, Ahn 2016[2]), using a dual Fabry-Pérot etalon, is a good example of this kind of instrument. The other way is to take a series of spectrograms using a grating-spectrograph with a fast scanning unit across the whole field of view (FOV). The greatest advantage for this kind of imaging spectroscopy is the capability to obtain primarily spectral data, which is the best for spectral studies of high-speed features. Additionally, its wide free spectral range contains both the strong lines and weak lines as well as the continuum. Thus, this method is useful for observing a target with different lines that are formed in different layers of the solar atmosphere. Moreover, as the entire spectrum is observed at the same position on the Sun simultaneously, the co-alignment between each wavelength is straightforward and not as serious problem as with the first method. The Fast Imaging Solar Spectrograph (FISS, Chae 2013[3]) of the GST is an example of this kind of instrument, designed to study the fine-scale structure and dynamics of chromospheric plasma, working in the visible wavelength.

To gain the advantages of the grating spectrograph for IR solar observations, we have developed CYRA as a new focal plane instrument for the GST. In this paper, we describe its design, control, data processing and sample result from the early runs.

## 2. CYRA DEVELOPMENT STRATEGY AND EXPECTED PERFORMANCE

The GST is configured as an unobstructed Gregorian system consisting of an off-axis parabolic primary, prime focus field stop and heat rejection reflector (heat-stop), off-axis elliptical secondary and diagonal flat mirrors. The telescope is designed with an equatorial mount and 1.6-meter clear aperture primary mirror. CYRA is located one floor beneath the telescope deck feed from the Gregorian focal plane. To achieve the scientific and instrumentation advantages of IR solar physics with CYRA, we need to acquire observational data with high temporal and high spatial resolution. The whole system consists of a Fore-Optics module plus internal elements. Table 1 lists preliminary instrumental characteristics of CYRA.

### 2.1 Fore-Optics Module

A set of mirrors located after the Gregorian focus relay the light down to the instrument floor. The slowly diverging beam after the Gregorian focus of the GST is collimated by a tilted spherical. This powered optics reimage the pupil onto a five-mirror image de-rotator. After the de-rotator, a set of two powered optics reimage the pupil onto a Tip/tilt mirror. A beam-splitter (B/S) located after this mirror reflects some of the light onto a context imager to provide observation live image, as well as a correlation tracker (CT) subsystem to measure in real time the image motion and apply a correction to the T/T mirror. The light transmitted through the B/S falls onto a powered optics which creates an image located just after an active image scanner. A field stop is positioned in the focal plane to create an intermediate slit. Finally, two more powered optics de-magnify and reimage the intermediate slit onto the real slit, inside the cryostat. An optional beam modulator system with polarizer and a rotating retarder located before the slit can provide polarized light for magnetic polarity measurement.

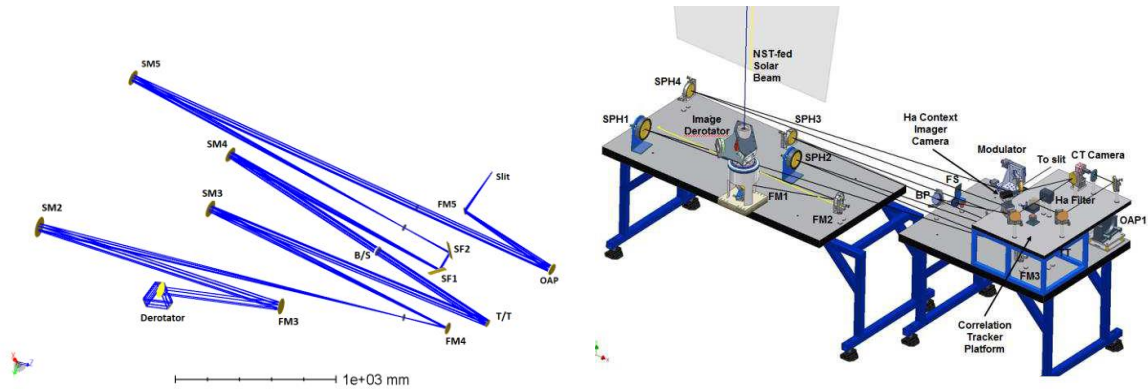

Figure 1. Optical path and illusionary capture for the Fore-Optics Module.

**Image De-rotator.** To correct the field rotation that results from telescope equatorial mount, a five-mirror de-rotation system has been designed by BBSO engineer team. This image de-rotator mounted on a stepping motor driven rotation stage that can orient the image to any arbitrary desired rotation angle and actively maintain the orientation during observations. The choice of five-mirror versus a more conventional three-mirror de-rotator is to minimize the angle of incidence on the mirrors and therefore, minimize the change of polarization as the de-rotator moves.

**Correlation Tracker.** Atmospheric seeing distortion as well as telescope tracking and flexure errors contribute to random FOV motions. The CT system consists of a T/T mirror and monitor camera. A reference image is taken by the monitor camera when the tracking process starts. Displacements between the live image and reference image, computed by the Sum of Absolute Difference (SAD) method, are converted to a voltage signal that is sent to the T/T mirror as feedback. By controlling the mirror tilting angle, the CT system will compensate the image displacement caused by atmospheric distortion and telescope tracking errors. The correction rate is about 400 Hz. This CT sub-system is developed by cooperators from the Korea Astronomy and Space Science Institute.

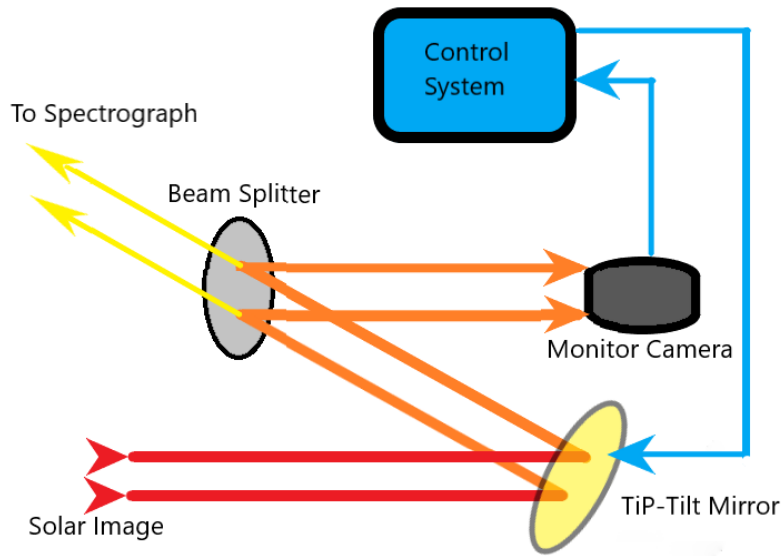

Figure 2. A diagram for context imager and CT system.

**Image Scanner.** The image scanner is a device that shifts the incident beam across the pre-slit for 2D imaging. The fold mirrors ride on a stepping motor translation stage as depicted in Figure 3. Each displacement of each step (8 μm or 16 μm) takes 70 ms with the positional error within 1 μm. The incoming beam is shifted by the mirrors perpendicular to the pre-slit. Only the part of the beam that pass through the pre-slit is sent to the internal elements.

In the default raster scan mode, the image scanner moves step by step, waits for the camera to record several frames before moving another step. The duration for each full scan is influenced by the number of frames taken at each step and the total step number, which is defined by the FOV and scan step density.

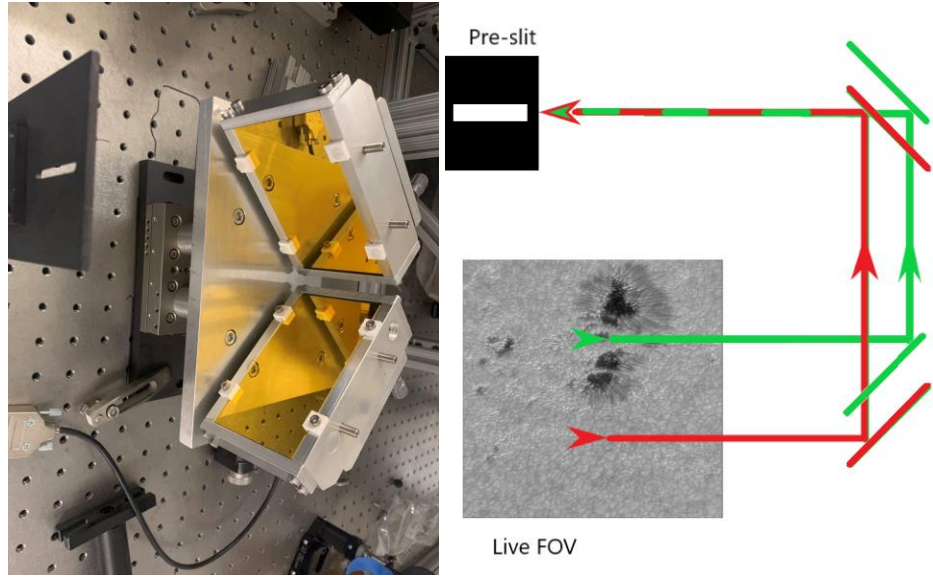

Figure 3. A diagram for the image scanner and pre-slit.

**Context Imager.** In the same arm as the CT, an Hα context imager provides a live image for the whole FOV. The positional precision of each raster scan step is verified by monitoring the simultaneous context image.

**Beam Modulator.** One of the scientific objectives for CYRA is to measure the vector magnetic field with magnetic sensitive lines in the infrared (e.g. Ti I triplet at 2.231 μm). Changing the single fold mirror FM5 to a beam-splitter, the instrument is switched from spectrograph to a dual-beam polarimeter mode. Currently a single beam modulator and linear polarizer has been assembled and tested for. The beam modulator and polarizer are mounted on a slide rail as an optional component for magnetic field measurement. During intensity observation, the modulator and linear polarizer are moved away from the beam to avoid energy loss. A dual beam modulator will be designed for CYRA in the future to increase the signal to noise ratio for polarization observation.

### 2.2 Internal Elements

All the internal elements, shown in Figure 4, are enclosed in a dual layer cryostat, which is kept cryogenic during observations (about 70 K for the outer chassis and 30 K for the inner chassis) to greatly decrease background infrared emission. The grating is 220 mm × 420 mm with a grating constant of 31.6 μm. Immediately inside the cryostat input window is a 10 positions filter wheel, which contains filters for different wavelengths each matched with an image slit of the appropriate width.

The grating is on a motorized rotation mount to allow selection of the spectral region and grating order from outside the cryostat. The spectral resolution ($R_0$) ranges from 150,000 up to 750,000 depending on wavelength. The final image is 9 mm in the spatial direction and 18 mm in the spectral. The image covers 500 vertical pixels for a plate scale of 0″.16/pixel, matching the GST's spatial resolution at a wavelength of 2 μm. Table 1 describes the parameter setup for the internal elements.

The spectrograph is a fully cryogenic folded Czerny-Turner optical design utilizing a 2048 × 2048 HgCdTe array with a quantum efficiency approaching 80% and a maximum frame rate of 76 Hz. The camera is able to work in three frequency modes -- 20 MHz, 40 MHz, and 80 MHz mode. The system is more stable in the high frequency mode but it provides longer exposure time in the low frequency mode.

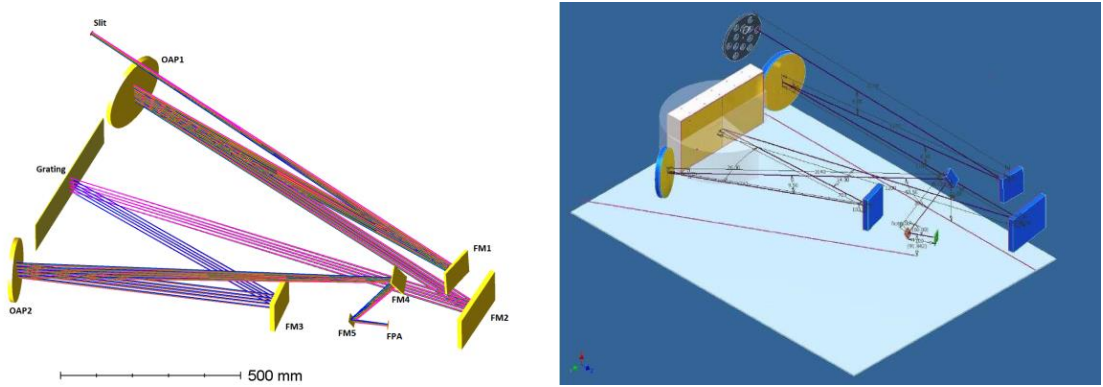

Figure 4. A diagram for internal elements.

Table 1. Parameter setup for the internal elements.

| Instrument parameters | Variable | Value |
| --- | --- | --- |
| Wavelength range | $\lambda$ | 2 - 5 μm |
| Field of view | FOV | 80″ × 80″ |
| Slit width | $\omega$ | 15 $\lambda$ |
| Image focal length | f | 1440 mm |
| Grating groove density | $1/\sigma$ | 31.6 mm$^{-1}$ |
| Grating blaze angle | $\varphi$ | 71° |
| Spectral resolving power | R | ~ $10^5$ |
| FPI | HAWWII II | HgCdTe |
| Camera format | $N_x \times N_y$ | 2048 × 2048 |
| Pixel size | $\delta_x \times \delta_y$ | 18 μm × 18 μm |

## 2.3 Control Software

Separate software systems were developed for the CT control and spectrograph observation.

**CT control software** consists calibration and tracking functions. Dark and flat filed corrections are applied to the live image. A sub-FOV region is selected to calculate two-dimensional displacement between the live image and reference image. The reference image can be periodically updated automatically to compensate for the ever-evolving features in the FOV. The control parameters in can be customized as needed for varying seeing conditions.

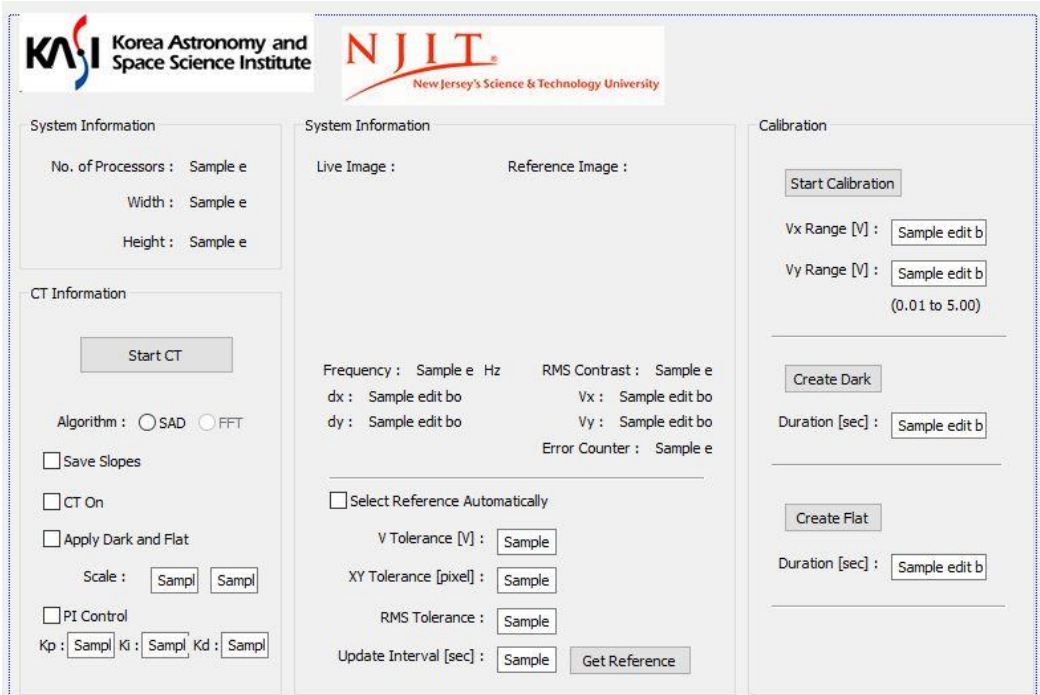

Figure 5. CT control software.

**CYRA observation software** is designed as an integrated control center for system calibrations and observation operations. The camera frequency can be switched between three modes to provide different exposure time to satisfy the corresponding working wavelength.

In general, the control software is designed to work in either monitor mode or observing mode. Device and software configuration should be applied in the monitor mode with live solar IR spectrum in the display window. The display range is customizable in two dimensions, and only data in the displayed region will be recorded during the observation to minimize the data rate.

Reference positions for each device has been remarked in the software interface. After the best setup for the solar spectrum has been confirmed, observers will start to collect spectrum data with the observing mode. Information of system configurations will be written in the header each spectral scan fits file.

Three types of spectral data can be acquired in the observing mode: standing spectrum, scanning spectrum and spectropolarimetric data.

  a) Standing spectrum. With the image scanner staying at a selected position, CYRA can achieve spectrum in its highest temporal cadence.
  b) Scanning spectrum. The image scanner will shift the FOV one step by one step at high precision to allow 2D spectral imaging. Scan range and step size is selected by the observers.
  c) Spectropolarimetric data. With the modulator and linear polarizer moved into the beam, CYRA can observe full Stokes (I/Q/U/V) date at each scanner position.

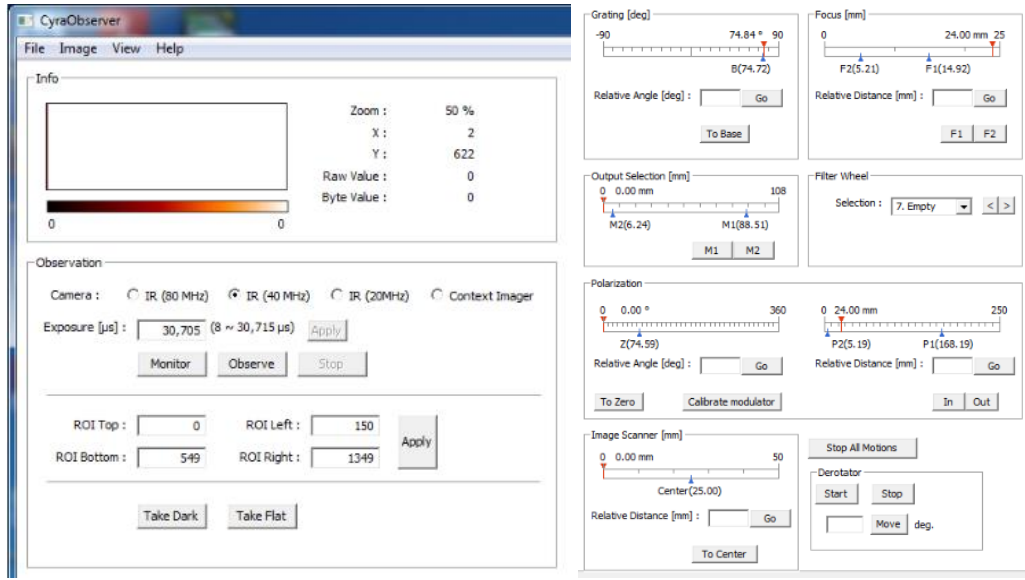

Figure 6. Control functions in CYRA observation software.

Table 2. Parameter for Ti band and CO band.

| Parameter | Ti Band | CO Band |
| --- | --- | --- |
| Filter central wavelength | 2231.37 nm | 4668.66 nm |
| Filter bandpass | 2225 nm -2238 nm | 4645 nm – 4693 nm |
| Filter transparency | 70% | 68% |
| Spectral order | 27th | 13th |
| Grating angle | 73.5° | 74.2° |
| Spectral resolution | 0.241 mÅ/pixel | 0.293 mÅ/pixel |
| Spectral resolving power | $3.84 \times 10^5$ | $1.67 \times 10^5$ |

## 3. DATA ACQUISITION AND REDUCTION

### 3.1 Data Acquisition and Reduction

CYRA can work in either spectrographic or spectropolarimetric mode. For each observation mode, three kinds of data, including science data, dark and flat field images, are recorded by the instrument. To obtain high quality scientific spectrum, certain data reduction processes should be applied to the raw data to remove instrumental effects.

Firstly, dark field data are taken by exposing with the telescope mirror cover is closed. The dark field thus consists of background signal that contains hot environment noise from the Fore-Optics. One hundred frames of dark images are recorded with the same exposure time as used for science data. The average of the dark fields are used for locating bad pixels and creating the dark image.

Secondly, flat fields data are taken several times during a day for each observing wavelength. This is accomplished by wobbling the telescope FOV around quiet sun region near solar disk center. An average frame of 1000 images is used to calibrate the spectrum slat and distortion, after subtracting the dark field and replacing bad pixels detected in the dark field. Thirdly, the non-uniform illuminations of spectrum will be corrected to obtain the flat field image. We decompose the flat pattern into the spatial pattern along the slit direction and the spectral pattern along the wavelength direction. The method is described by Chae 2013[3].

After dark and flat field corrections, residual bad pixels and hot pixels will be removed from the science data by using a list created from the dark field. To obtain the slit tilt information, we placed a thin wire target at the pre-slit and calculate the spectral slant. By simulating the telluric line core position, the solar spectrum curvature can be corrected.

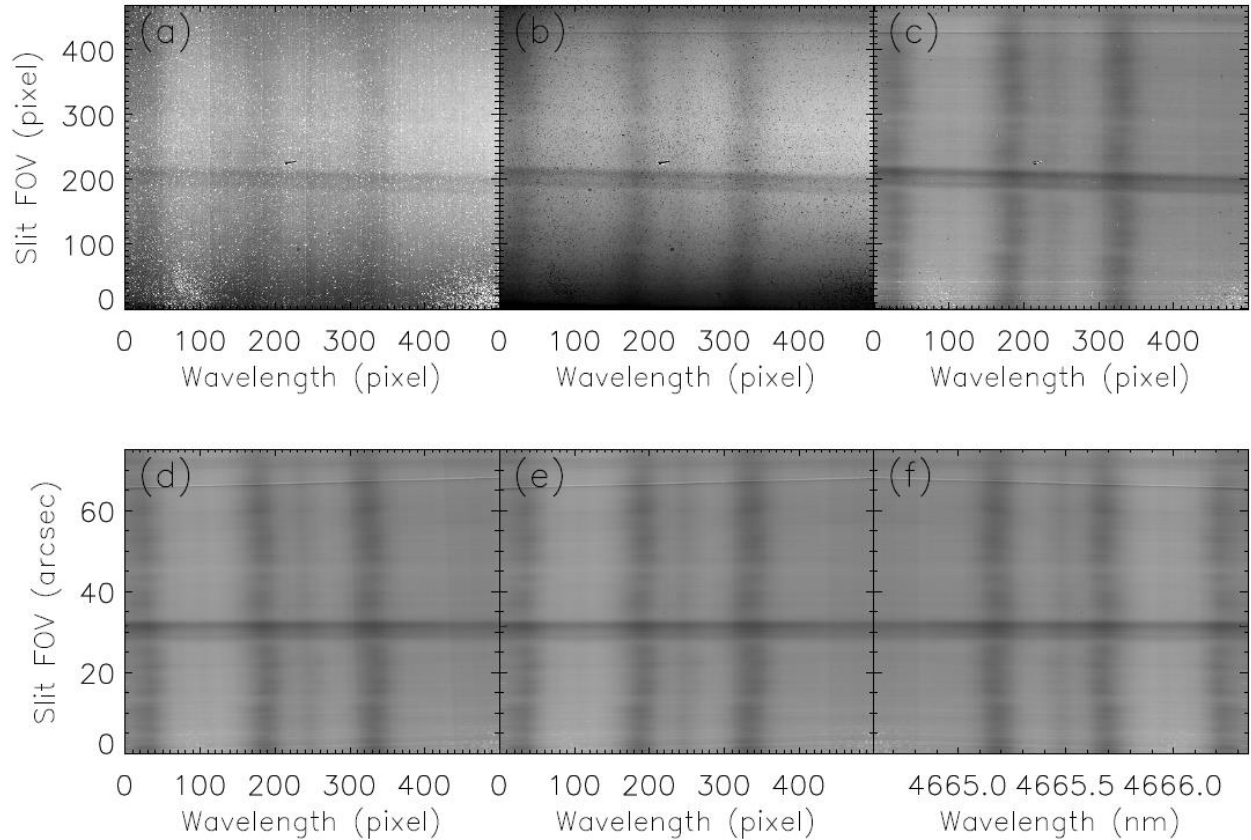

Figure 7. Data reduction for CO lines near 4667 nm. The raw data (panel a) is processed with dark fielding (panel b), flat fielding (panel c), bad pixel and hot pixel correction (panel d) and slit slant as well as spectral line tilt/curvature correction to get usable spectrum as in panel e. From panel e to panel f, the spectrum has been flipped in the wavelength direction and the unit for the x-axis has been calibrated from pixel to physical wavelength.

**3.2 Testing Observations**

We have had several testing observations for CYRA. The observing pipeline is described as follows:

a. Direct the sunlight to the Coudé room to confirm the telescope system is focusing well, then switch the beam to CYRA.

b. Point the telescope at the polar limb of the Sun to calibrate the image orientation with the live image, then enable image de-rotator. Point to the target and take dark and flat field for CT at the nearby Quiet Sun region. Close loop for the CT on a selected feature.

c. Check the initial position of the image scanner and let center of the FOV travel through the pre-slit. Choose the appropriate filter for the observing wavelength. Locate the target spectrum in the display window.

d. For spectrograph mode, switch the beam modulator off from the beam and remove the linear polarizer. For spectropolarimeter mode, switch the beam modulator into the beam and add the linear polarizer.

e. Click "Observe" button to setup scan range, scan step size and total number of scans and start to acquire science data. Every few hours, the sunlight should be sent back to the Coudé room to check the telescope focus. Dark and flat field data can be taken during this time.

With one burst of scanning spectroscopic data, researchers can produce an IR solar image by stacking plots of a selected wavelength from each spectrum. Figure 8 compares a sample sunspot near disk center in active region NOAA 12690 observed on 2017-09-15 by Solar Dynamics Observatory (SDO), Interface Region Imaging Spectrograph (IRIS) and CYRA. Spectral intensity, line width and Doppler velocity information were utilized to create maps for the active region. Comparing with the SDO and IRIS data, CYRA did not achieve its designed spatial resolution (0.32") of this observation. However, its spectrum performed well and resolved fine 3 minutes and 5 minutes oscillations in continuous observation.

The problem of low spatial resolution was a result of undersampling the raster scans and camera defocus. In another observation taken on 2018-09-11, we achieved expected high spatial resolution after fine tuning the optical system. Figure 9 compares the SDO imaging data and CYRA raster image. The structures of dark pores and bright network are clear. It is noteworthy that unusual absorption features near the pores and network are presenting in the CYRA CO Raster image, however they are not observed by SDO. As an expense of higher spatial resolution, the spectral resolution is not as high as in the previous observation. The problem maybe due to astigmatism including further tuning to the optical system is still needed.

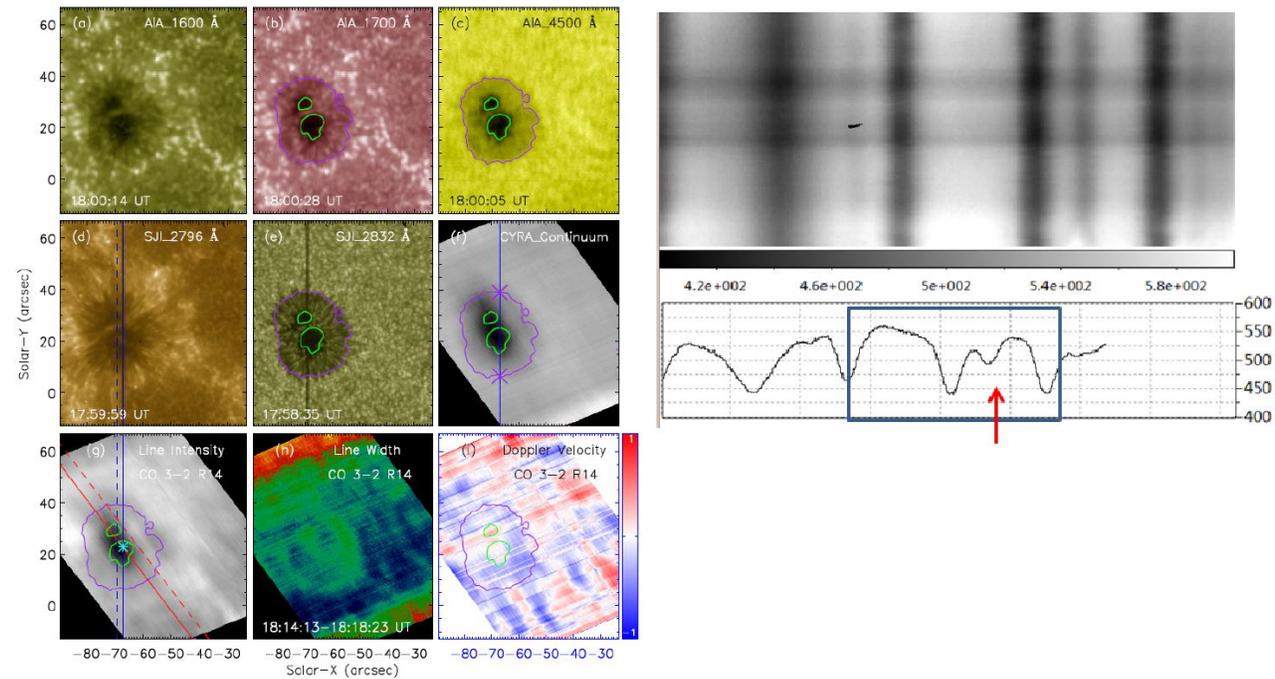

Figure 8. Observation to solar active region NOAA 12680 on 2017-09-15. Left: panel a, b and c are SDO imaging data. Panel d and e are IRIS raster images. Panel f to I are CYRA raster images for continuum intensity, CO 3-2 R14 spectral line intensity, line width and Doppler velocity, respectively. Right panel: Spectrum around 4665 nm observed by CYRA and a sample intensity profile.

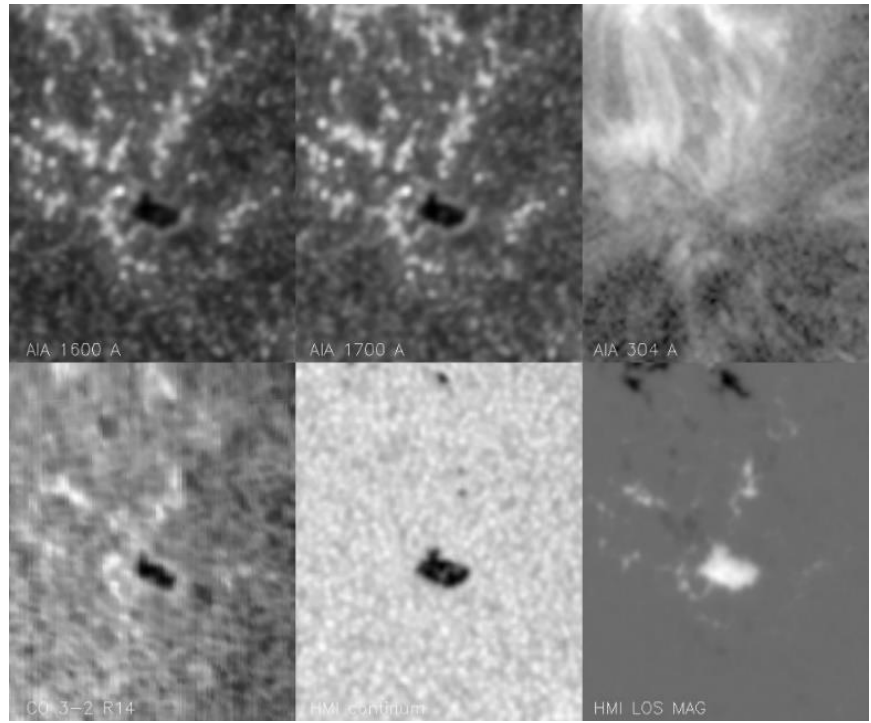

Figure 9. Observation to a pore in NOAA 12722 on 2018-09-11. The lower left panel presents CYRA raster image observed by CO 3-2 R14. Other panels are imaging data observed by SDO.

## 4. SUMMARY

We have described the design, specifics, data reduction, and sample result from the CYRA. The instrument was designed to take spectrum of features in chromosphere and temperature minimum at high spatial-, high spectral- and high temporal- resolution. However, it is not possible to meet all the requirements and it is necessary to compromise among different needs. It is proposed that further fine tuning of the optical system will enable the system to achieve these goals. Figure 8 and Figure 9 provide example data currently obtained by CYRA. This instrument satisfies the requirement to observe solar IR spectrum for scientific research. Observation with CO lines around 4667 nm and Ti I triplet around 2231 nm are the first targets.

We plan to improve CYRA in several stages. The first stage is to upgrade the CT to a low order adaptive-optics system to achieve better image stabilizing quality and capability to correct for static image aberration. The second stage will be to replace the current camera of the context imager to obtain better image quality. The final stage will be to replace the IR array detector now in use with a new science grade IR array to improve data quality with increased signal to noise ratio.

**Acknowledgements.** BBSO operation is supported by NJIT and US NSF AGS-1821294 grant. GST operation is partly supported by the Korea Astronomy and Space Science Institute, the Seoul National University, and the Key Laboratory of Solar Activities of Chinese Academy of Sciences (CAS) and the Operation, Maintenance and Upgrading Fund of CAS for Astronomical Telescopes and Facility Instruments.

## REFERENCES

[1] Cao, W., Gorceix, N., Coulter, R., Ahn, K., Rimmele T.R., Goode, Philip R., "Scientific instrumentation for the 1.6 m New Solar Telescope in Big Bear, " Astronomische Nachrichten, 331, 6, 636

[2] Ahn, Kwangsu., Cao, Wenda., Shumko, Sergiy., Chae, Jongchul., "Data Processing of the magnetograms for the Near InfraRed Imaging Spectropolarimeter at Big Bear Solar Observatory," American Astronomical Society, SPD meeting #47, id.2.07

[3] Chae, Jongchul., Park, Hyung-Min., Ahn, Kwangsu., Yang, Heesu., Park, Young-Deuk., Nah, Jakyoung., Jang, Bi Ho., Cho, Kyung-Suk., Cao, Wenda., Goode, Philip R.,"Fast Imaging Solar Spectrograph of the 1.6 Meter New Solar Telescope at Big Bear Solar Observatory," SoPh. 228, 1, 1-22 (2013).